\documentclass[journal]{IEEEtran}

\usepackage{amsmath,amsthm,amssymb,amsfonts}
\usepackage{graphicx}
\usepackage{epstopdf}

\usepackage{array,booktabs,multirow,makecell,tabularx,tikz}
\usepackage{colortbl}

\usepackage[linesnumbered, ruled, boxed]{algorithm2e}
\usepackage{algpseudocode}

\usepackage{caption,subcaption}
\usepackage{fancyhdr}
\usepackage{float}
\usepackage{stmaryrd}
\usepackage{dsfont}
\usepackage{color}

\makeatletter
\def\ps@plain{%
  \let\@mkboth\@gobbletwo%
  \let\@oddhead\@empty%
  \let\@evenhead\@empty%
  \let\@oddfoot\@empty%
  \let\@evenfoot\@empty%
}

\def\ps@headings{%
  \let\@mkboth\markboth%
  \let\@oddhead\@empty%
  \let\@evenhead\@empty%
  \let\@oddfoot\@empty%
  \let\@evenfoot\@empty%
}

\def\ps@IEEEtitlepagestyle{%
  \let\@mkboth\@gobbletwo%
  \let\@oddhead\@empty%
  \let\@evenhead\@empty%
  \let\@oddfoot\@empty%
  \let\@evenfoot\@empty%
}
\makeatother
\allowdisplaybreaks[4]

\definecolor{headercolor}{RGB}{52, 101, 164}
\definecolor{rowcolor1}{RGB}{242, 246, 252}
\definecolor{rowcolor2}{RGB}{255, 255, 255}

\newtheorem{theorem}{Theorem}

\newtheorem{lemma}{Lemma}

\newtheorem{proposition}{Proposition}

\newtheorem{corollary}{Corollary}

\newtheorem{property}{Property}

\newtheorem{remark}{Remark}

\newtheorem{claim}{Claim}

\newcolumntype{A}{>{\hsize=0.95\hsize\raggedright\arraybackslash}X}  % Component
\newcolumntype{B}{>{\hsize=0.6\hsize\centering\arraybackslash}X}   % Type
\newcolumntype{C}{>{\hsize=1.7\hsize\centering\arraybackslash}X}   % In out
\newcolumntype{D}{>{\hsize=0.75\hsize\centering\arraybackslash}X}  % Activation

\SetKwProg{Init}{Initialization}{}

\usepackage{hyperref}
\hypersetup{
  colorlinks=false,
  urlcolor=black,
  linkcolor=black,
  citecolor=black,
  pdfborder={0 0 0}
}

\begin{document}
 
 \pagestyle{empty}

\title{ \huge Group Relative Policy Optimization for Robust \\ Blind Interference Alignment with Fluid Antennas
}

\author{
Jianqiu Peng, Tong Zhang, Shuai Wang, Mingjie Shao, Hao Xu, and Rui Wang \vspace{-0.20cm}
 
\thanks{J. Peng is with the School of Information Science and Technology, Harbin Institute of Technology, Shenzhen, China (email: 25s152083@stu.hit.edu.cn).}

\thanks{T. Zhang is with Guangdong Provincial Key Laboratory of Aerospace Communication and Networking Technology, Harbin Institute of Technology, Shenzhen, 518055, China (e-mail: tongzhang@hit.edu.cn).}

\thanks{S. Wang is with the Shenzhen Institute of Advanced Technology, Chinese Academy of Sciences, Shenzhen 518055, China (e-mail: s.wang@siat.ac.cn).}

\thanks{M. Shao is with the KeyLaboratory of Systems and Control, Institute of Systems Science, Academy of Mathematics and Systems Science (AMSS), Chinese Academy of Sciences (CAS), Beijing 100149, China (e-mail: mingjieshao@amss.ac.cn).}

\thanks{\textcolor{black}{H. Xu is with the School of Information Science and Engineering, Southeast University, Nanjing, 210096, P.R. China (e-mail: hao.xu@seu.edu.cn).}}

\thanks{R. Wang are with the Southern University of Science and Technology (e-mail: wangr@sustech.edu.cn).}

\thanks{This work was supported by
the National Natural Science Foundation of China under Grants 62401229, 62501152, 62401340, U25A20398; by the Shenzhen Science and Technology Program under Grants JCYJ20241202124934046,  RCYX20231211090206005; by the Fundamental Research Funds for the Central Universities under Grant 2242025R10001.}

\thanks{Our code is available at {\color{cyan}\url{https://github.com/JianqiuPeng/GRPO_CUDA}}.}

}

% 添加紧凑间距命令
\makeatletter
\def\IEEEtitleabstractindextextspace{\vspace{-0.5\baselineskip}}
\makeatother

\maketitle

\begin{abstract}
Fluid antenna system (FAS) leverages dynamic reconfigurability to unlock spatial degrees of freedom and reshape wireless channels. Blind interference alignment (BIA) aligns interference through antenna switching. This paper proposes, for the first time, a robust fluid antenna-driven BIA framework for a $K$-user MISO downlink under imperfect channel state information (CSI). We formulate a robust sum-rate maximization problem through optimizing fluid antenna positions (switching positions). To solve this challenging non-convex problem, we employ group relative policy optimization (GRPO), a novel deep reinforcement learning algorithm that eliminates the critic network. This robust design reduces model size and floating point operations (FLOPs) by nearly half compared to proximal policy optimization (PPO) while significantly enhancing performance through group-based exploration that escapes bad local optima. Simulation results demonstrate that GRPO outperforms PPO by $4.17\%$, and a $100$K-step pre-trained PPO by $30.29\%$. Due to error distribution learning, GRPO exceeds heuristic MaximumGain and RandomGain by $200.78\%$ and $465.38\%$, respectively.
\end{abstract}


% 确保内容立即开始
%\vspace{-0.2cm}

\section{Introduction}

Fluid antenna systems (FAS) exploit dynamic reconfigurability to unlock spatial degrees of freedom (DoF) and reshape wireless channels. We introduce fluid antenna-driven blind interference alignment (BIA). This paper maximizes sum-rate in BIA system under channel estimation errors using deep reinforcement learning (DRL), specifically group relative policy optimization (GRPO). 

FAS was seminally introduced to wireless communications by Wong et al. in \cite{FAS}. Then, FAS was widely investigated in various fields and shows significant improvement due to its assistance, including integrated sensing and communication (ISAC) \cite{Qian}, symbol-level precoding \cite{Wenxuan}, smart radio environment \cite{SRE-FAS}, etc. In addition to the fields above, multiple access also plays a vital role in wireless communications, since it is generally desirable to serve more users on the same radio resource. Recent research shows that FAS significantly enhances multiple access through their distinctive fluid capacity \cite{MAC-Hao-1,MAC-Hao-2, MAC-Slow, MAC-Mahdi, MAC-DL, MAC-Waqar}, so-called fluid antenna multiple access (FAMA). In particular, \cite{MAC-Hao-1} considered a FAMA system. By dynamically selecting spatial nulls of deep fading for each user, FAS converts multi-user interference into capacity gains, significantly enhancing multiple access channel capacity. \cite{MAC-Hao-2} investigated the outage probability of a two-user FAMA system, employing more accurate spatial correlation modeling and analysis to reveal how the number of antenna ports and their size impact system performance, and proposed low-complexity approximation methods. \cite{MAC-Slow} proposed the slow FAMA, where each user is equipped with a fluid antenna with switchable ports that select the port with the optimal signal-to-interference ratio (SIR) when channel conditions change, thereby suppressing multi-user interference.

A  critical direction of FAMA is to minimize the reliance on channel state information at the transmitter (CSIT), or even eliminate it. In particular, with a reduced amount of CSIT, \cite{MAC-Mahdi} proposed a slow FAMA scheme based on conditional generative adversarial networks, which can generate the signal-to-interference-ratio (SINR) distribution of all ports by observing only a small number of port SINRs. In addition, \cite{MAC-DL} proposed a port selection scheme for slow FAMA systems based on a long short-term memory (LSTM) deep learning model, which can infer the SINR at all ports using only a small number of port observations. Without CSIT, \cite{MAC-Waqar} proposed the ``Turbo FAMA" method, which significantly enhances the interference suppression capability and user capacity of FAMA by integrating heuristic port pre-selection, maximum ratio combining, deep joint source-channel coding, and a hybrid diffusion-based denoising model. However, a critical gap remains: all existing studies on CSIT-reduced FAMA is solely concerned with fast fading channels, leaving its implementation in static environments without CSIT entirely unexplored.

%\textit{``How to achieve multiple-access for FAS with no CSIT?"}
Fortunately, we can bridge this gap with blind information alignment (BIA), a well-established concept from information theory \cite{BIA-3,BIA-1,BIA-2,BIA-4,BIA-5}. BIA achieves interference alignment without CSIT by switching the reconfigurable antenna modes. This creates temporal channel correlations that confine interference to a reduced-dimensional subspace at interfered receivers, while preserving the separability of desired signals at the desired receiver. \cite{BIA-1} designed a BIA transmission scheme with antenna switching patterns for the $K$-user multiple-input single-output (MISO) downlink. For the $K$-user interference channel,  BIA  schemes have been investigated in \cite{BIA-2} and \cite{BIA-5}, which demonstrate that the achievable DoF can scale linearly with the number of users $K$. \cite{BIA-4} proposed a BIA scheme for partially connected cellular networks. Although existing studies on BIA have laid important groundwork, key implementation issues such as sum-rate maximization and robust design have not been investigated.

\begin{figure*}[t]
\centering
\includegraphics[width=5.5in]{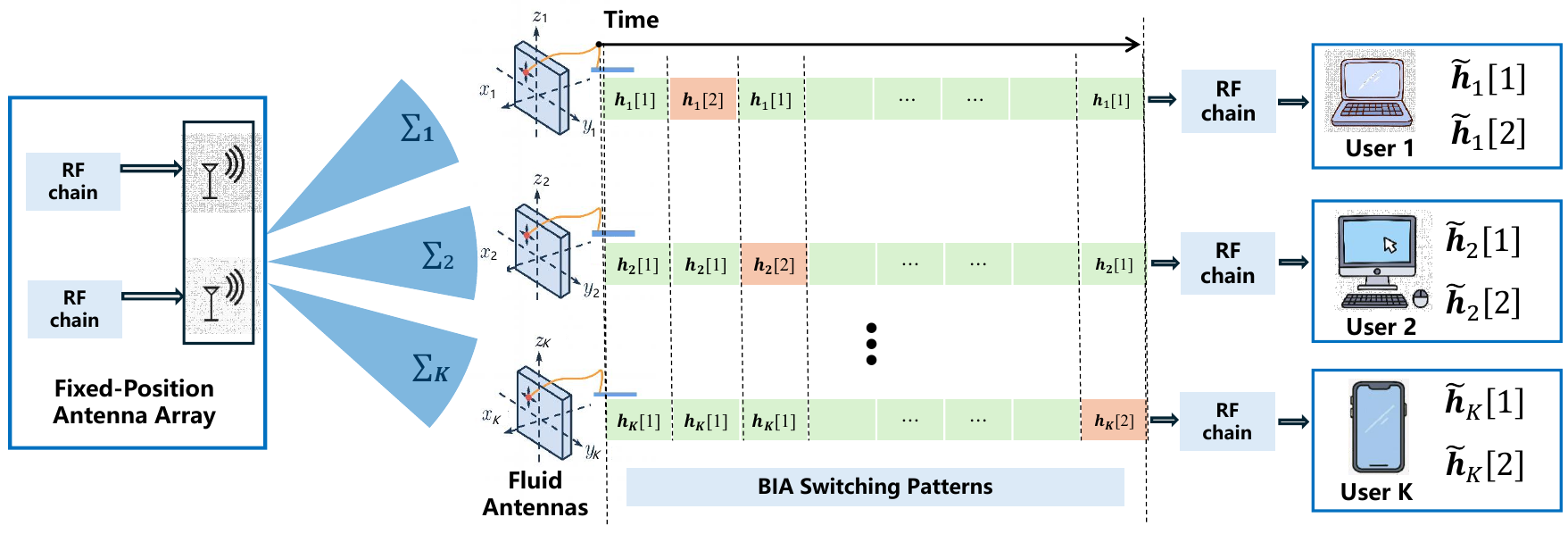}
\caption{Illustration of the considered $K$-User MISO downlink BIA system model}
\end{figure*}

In this paper, we investigate a fluid antenna-assisted $K$-user  MISO downlink system. A base station (BS) with two fixed antennas serves $K$ users, each equipped with a fluid antenna. Under imperfect channel state information (CSI), we adopt the BIA scheme from \cite{BIA-1} and formulate a robust sum-rate maximization problem. To solve this non-convex problem, we employ the GRPO algorithm \cite{GRPO}. By eliminating the critic network, GRPO not only reduces the model size and floating point operations (FLOPs)  by nearly half compared to proximal policy optimization (PPO) \cite{PPO}, but also enhances the rate. The removal of the critic resets the exploration space, enabling group-based exploration to escape bad local optima of PPO. On average across different CSI error, simulations show that the GRPO solution outperforms PPO by $4.17\%$. Due to error distribution learning, GRPO exceeds a $100$K-step PPO pre-training by $30.29\%$, and surpasses the heuristic MaximumGain and RandomGain by $200.78\%$ and $465.38\%$, respectively.

%\textit{Notations}: 

\section{$K$-User MISO Downlink BIA System Model and Problem Formulation}

As shown in Fig. 1, we consider a $K$-user MISO downlink BIA system, where the BS is equipped with a fixed-position antenna array with $2$ antennas, and each user is equipped with a fluid antenna. There are $K$ users and the BS aims to simultaneously transmit data to all users.  We assume no CSIT, while perfect CSI is available at the users. Furthermore, the wireless channels are assumed to be static, with all temporal variations artificially induced solely by the dynamic movement of fluid antennas. The time is slotted and the input-output relationship at user $k\, \in \textcolor{black}{\{1,\cdots,K\}}$ in the time slot $t\,\in \textcolor{black}{\{1,\cdots,K\}}$ is given as follows: 
\begin{equation}
    y_k(t) = \mathbf{h}_k(t)\sum_{k=1}^K\sqrt{P}\mathbf{s}_k(t)  + n_k(t), 
\end{equation}
where $y_k(t)\,\in \mathbb{C}$ denotes the received signal at the user $k$ and in the time slot $t$; $\mathbf{h}_k(t)\,\in \mathbb{C}^{1 \times 2}$ denotes the channel from BS to \textcolor{black}{the user $k$ in the time slot $t$}; $P$ denotes the transmit power; $\mathbf{s}_k(t)$ denotes transmit symbols for \textcolor{black}{the user $k$ in the time slot $t$} satisfying $\mathbb{E}\{\|\mathbf{s}_k(t)\|^2\} = 1$; and $n_k(t) \sim \mathcal{CN}(0,\sigma^2)$ denotes the additive \textcolor{black}{white} Gaussian noise (AWGN) \textcolor{black}{at the user $k$  in the time slot $t$}.  

\subsubsection{Field Response Channel Model}  
Based on the field response channel model \cite{Xiao}, we assume that the fluid antenna of user $k$ is located at position $\mathbf{u}_{k,n} = (x_k[n], y_k[n], z_k[n]),\,\textcolor{black}{n \in \{1,2\}}$. Then, the channel can be expressed as
\begin{equation}
\mathbf{h}_k[n] = \mathbf{f}_k^H(\mathbf{u}_{k,n}) {\bf \Sigma}_k \mathbf{G}_k,
\end{equation}
where $\mathbf{f}_k(\mathbf{u}_{k,n}) = [e^{j \frac{2\pi}{\lambda} \rho_1^r(\mathbf{u}_{k,n})}; \dots; e^{j \frac{2\pi}{\lambda} \rho_{L_k^r}^r(\mathbf{u}_{k,n})} ] \in \mathbb{C}^{L_k^r \times 1}$ denotes the receive field response vector (FRV); ${\bf{\Sigma}}_k \in \mathbb{C}^{L_k^r \times L_k^t}$ is the path-response matrix (PRM), whose element $[{\bf {\Sigma}}_k]_{j,i}$ represents the channel response coefficient between the  transmit path $i \in \textcolor{black}{\{1,\cdots,L_k^t\}}$ and the receive path $j \in \textcolor{black}{\{1,\cdots,L_k^r\}}$ for the user $k$; $\mathbf{G}_k = [\mathbf{g}_k(\mathbf{v}_1), \mathbf{g}_k(\mathbf{v}_2)] \in \mathbb{C}^{L_k^t \times 2}$ is the transmit field response matrix (FRM), with each column vector $\mathbf{g}_k(\mathbf{v}_m) = [e^{j \frac{2\pi}{\lambda} \rho_1^t(\mathbf{v}_m)}; \dots; e^{j \frac{2\pi}{\lambda} \rho_{L_k^t}^t(\mathbf{v}_m)} ] \in \mathbb{C}^{L_k^t \times 1}$ corresponding to the fixed position transmit antenna $\textcolor{black}{m \in \{1,2\}}$. Specifically, the phase term $\rho^r_j(\mathbf{u}_{k,n}) = x^r_{k,n} \eta^r_j + y^r_{k,n} \beta^r_j + z^r_{k,n} \omega^r_j$ corresponds to the receive path $j \in \{1,\cdots,L_k^r\}$, where the direction parameters are given in terms of the pitch angle $\theta^r_j \in [-\pi/2,\pi/2]$ and azimuth angle $\phi^r_j \in [-\pi/2,\pi/2]$ as follows:
\begin{equation}
\eta^r_j = \cos \theta^r_j \cos \phi^r_j, \,\,
\beta^r_j = \cos \theta^r_j \sin \phi^r_j, \,\,
\omega^r_j = \sin \theta^r_j.
\end{equation}
Similarly, the phase term $\rho^t_i(\textbf{v}_m) = x_m^t \eta^t_i + y_m^t \beta^t_i + z_m^t \omega^t_i$ corresponds to the transmit path $i \in \{1,\cdots,L_k^t\}$, with pitch angle $\theta^t_i \in [-\pi/2,\pi/2]$ and azimuth angle $\phi^t_i \in [-\pi/2,\pi/2]$, and the associated parameters are:
\begin{equation}
\eta^t_i = \cos \theta^t_i \cos \phi^t_i, \,\,
\beta^t_i = \cos \theta^t_i \sin \phi^t_i, \,\,
\omega^t_i = \sin \theta^t_i.
\end{equation}

\subsubsection{Imperfect CSI Model} \textcolor{black}{Due to} channel estimation errors or the time shift of the channel, we assume that the CSI at the users are imperfect. In particular, imperfection is related to the estimated channel patterns, that is, 
\begin{equation}
    \mathbf{h}_k[m] = \widetilde{\mathbf{h}}_k[m] + {\bf{\Delta}}_k[m],  
\end{equation}
where according to \cite{ZhouGui}, channel estimation error ${\bf{\Delta}}_k[m] \in \mathbb{C}^{1 \times 2}$ for channel pattern $m$ at the user $k$ can follow a complex Gaussian distribution $\mathcal{CN}(\mathbf{0},\mathbf{V}_k[m])$ with a semi-definite covariance matrix $\mathbf{V}_k[m]$; $\mathbf{h}_k[m]$ denotes the real channel pattern $m$ at the user $k$; and $\widetilde{\mathbf{h}}_k[m]$ denotes the estimated channel pattern $m$ at the user $k$.

\subsection{BIA Scheme in \cite{BIA-1} for Two Transmit Antennas}

\subsubsection{Two-User Example in \cite{BIA-1}}

For this case, the achievable DoF is $4/3$, namely each user \textcolor{black}{can transmit} $2$ data symbols over $3$ time slots. \textcolor{black}{In} the time slot $1$, BS transmits:
\begin{equation}
    \mathbf{x}(1) = \sqrt{P}(\mathbf{s}_1 +  \mathbf{s}_2),
\end{equation}
where $\mathbf{s}_1 \in \mathbb{C}^{2 \times 1}$ \textcolor{black}{contains} two symbols desired by user $1$ and $\mathbf{s}_2 \in \mathbb{C}^{2 \times 1}$ \textcolor{black}{contains} two symbols desired by user $2$. The received signals are given by
\begin{subequations}
    \begin{eqnarray}
&&    y_1(1) = \mathbf{h}_1[1]\sqrt{P}(\mathbf{s}_1 +  \mathbf{s}_2) + n_1(1),  \\
&&    y_2(1) = \mathbf{h}_2[1]\sqrt{P}(\mathbf{s}_1 +  \mathbf{s}_2) + n_2(1), 
\end{eqnarray}
\end{subequations}
where fluid antenna of the user $k$ is at ``Position $1$" resulting in $\mathbf{h}_k[1],\, k \in \{1,2\}$. \textcolor{black}{In} the time slot $2$, BS transmits:
\begin{equation}
    \mathbf{x}(2) = \sqrt{P}\mathbf{s}_1. 
\end{equation}
The received signals are given by
\begin{subequations}
    \begin{eqnarray}
&&    y_1(2) = \sqrt{P}\mathbf{h}_1[2]\mathbf{s}_1 + n_1(2),  \\
&&    y_2(2) = \sqrt{P}\mathbf{h}_2[1]\mathbf{s}_1 + n_2(2), 
\end{eqnarray}
\end{subequations}
where fluid antenna of the user $1$ is at ``Position $2$" resulting in $\mathbf{h}_1[2]$, and the fluid antenna of user 2 is at ``Position $1$" resulting in $\mathbf{h}_2[1]$. \textcolor{black}{In} the time slot $3$, BS transmits:
\begin{equation}
    \mathbf{x}(3) = \sqrt{P}\mathbf{s}_2. 
\end{equation}
The received signals are given by
\begin{subequations}
    \begin{eqnarray}
&&    y_1(3) = \sqrt{P}\mathbf{h}_1[1]\mathbf{s}_2 + n_1(3),  \\
&&    y_2(3) = \sqrt{P}\mathbf{h}_2[2]\mathbf{s}_2 + n_2(3), 
\end{eqnarray}
\end{subequations}
where fluid antenna of the user $1$ is at ``Position $1$" resulting in $\mathbf{h}_1[2]$, and the fluid antenna of the user 2 is at ``Position $2$" resulting in $\mathbf{h}_2[2]$. 
Note that ``Position $1$" and ``Position $2$" may correspond to different fluid antenna locations for different users. For decoding, user $1$ decodes $\mathbf{s}_1$ from
\begin{eqnarray}
    \begin{bmatrix}
        y_1(1) - y_1(3)  \\
        y_1(2)
    \end{bmatrix} =  \sqrt{P}\begin{bmatrix}
  \widetilde{\mathbf{h}}_1[1] \\
 \widetilde{\mathbf{h}}_1[2]
    \end{bmatrix} \mathbf{s}_1 +  \sqrt{P}\begin{bmatrix}
 {\bf{\Delta}}_1[1]   \\
  {\bf{\Delta}}_1[2] 
    \end{bmatrix} \mathbf{s}_1
 \nonumber \\    + \begin{bmatrix}
   n_1(1) - n_1(3)  \\
        n_1(2)
     \end{bmatrix},
\end{eqnarray} and user 2 decodes $\mathbf{s}_2$ from 
\begin{eqnarray}
    \begin{bmatrix}
        y_2(1) - y_2(2)  \\
        y_2(3)
    \end{bmatrix} = \sqrt{P}\begin{bmatrix}
\widetilde{\mathbf{h}}_2[1] \\
\widetilde{\mathbf{h}}_2[2]
    \end{bmatrix} \mathbf{s}_2 + 
    \sqrt{P}\begin{bmatrix}
 {\bf{\Delta}}_2[1]   \\
  {\bf{\Delta}}_2[2] 
    \end{bmatrix} \mathbf{s}_2
    \nonumber \\  + \begin{bmatrix}
    n_2(1) - n_2(2)  \\
        n_2(3)
     \end{bmatrix}.
\end{eqnarray}

\subsubsection{$K$-User BIA Scheme in \cite{BIA-1} and Rate Expression}
We consider the scheme in \cite{BIA-1} for $2$ transmit antennas, which achieves $\frac{2K}{K+1}$  DoF. The scheme spans $K+1$ time slots, and each user decodes $2$ symbols. The procedure is described as follows: \textcolor{black}{In} the time slot $1$, the BS transmits a linear combination of symbols intended for all users, i.e., $\mathbf{s}_1 + \mathbf{s}_2 + \cdots$, while all users remain in ``Position $1$''. In the time slot $2$, the BS transmits only the symbol intended for user $1$. At this time, user $1$ switches to ``Position $2$'', while all other users remain in ``Position $1$''. This pattern continues similarly in the subsequent time slots. \textcolor{black}{In} the time slot $K+1$, BS transmits only the symbol for user $K$, and user $K$ switches to ``Position $2$'', with the remaining users still in ``Position $1$''. After interference subtraction, each user can decode two symbols via an equivalent point-to-point channel, given by
\begin{eqnarray}
    \begin{bmatrix}
        y_k(1) - \sum_{i=1,i\ne k}^K y_k(i+1) \\
        y_k(k+1)
    \end{bmatrix} = \sqrt{P}
    \underbrace{\begin{bmatrix} 
\widetilde{\mathbf{h}}_k[1] \\ 
\widetilde{\mathbf{h}}_k[2] 
    \end{bmatrix}}_{\text{denoted\,as\,} \widetilde{\mathbf{H}}_k} \mathbf{s}_k + \nonumber \\         \sqrt{P}\underbrace{\begin{bmatrix}
 {\bf{\Delta}}_k[1]   \\
  {\bf{\Delta}}_k[2] 
    \end{bmatrix}}_{\text{denoted\,as\,} {\bf{\Delta}}_k} \mathbf{s}_k  + \underbrace{\begin{bmatrix}
        n_k(1) - \sum_{i=1,i\ne k}^K n_k(i+1) \\
        n_k(k+1)
    \end{bmatrix}}_{\text{denoted\,as\,} \mathbf{n}_k}, \label{IO}
\end{eqnarray}
where $\mathbf{s}_k \in \mathbb{C}^{2 \times 1}$ denotes the symbols desired by the user $k$.  

\textbf{Proposition $1$}: For the $K$-user BIA scheme in \cite{BIA-1} with $2$ antennas, the rate expression for the user $k$ is given by 
\begin{equation}
    R_k = \log_2 |\mathbf{I}_2 + P \widetilde{\mathbf{H}}_k \widetilde{\mathbf{H}}_k^H \textcolor{black}{{\bf \Omega}}^{-1}|,
\end{equation}
where $\widetilde{\mathbf{H}}_k$ is defined in \eqref{IO} and 
\begin{equation}
\!\!\!\!\! \textcolor{black}{{\bf \Omega}} = P \begin{bmatrix}
     {\bf{\Delta}}_k[1] {\bf{\Delta}}_k^H[1] + K \sigma^2 & {\bf{\Delta}}_k[1] {\bf{\Delta}}_k^H[2] \\
     {\bf{\Delta}}_k[2] {\bf{\Delta}}_k^H[1] & {\bf{\Delta}}_k[2] {\bf{\Delta}}_k^H[2] + \sigma^2
 \end{bmatrix}. \label{Rn}
\end{equation}
\begin{IEEEproof}
    Please refer to Appendix A. 
\end{IEEEproof}

\subsection{Fluid Antenna Position Optimization with Imperfect CSI}

Our objective is to maximize the sum-rate via optimizing the fluid antenna positions (switching positions). Mathematically, we formulate the problem as follows:
\begin{subequations}
\begin{align}
(\text{P1})\ \max_{\{\mathbf{u}_{k,n}\}} \ & 
\sum_{k=1}^K  R_k \\
\text{s.t.} \ 
& \|\mathbf{u}_{k,n} - \mathbf{u}_{k,n'}\|_2 \geq d_{\min}, \ \forall k, n, n' \label{C1} \\
& \mathbf{u}_{k,n} \in \mathcal{C}_k, \qquad \qquad \qquad \,\, \forall k, n \label{C2}
\end{align}
\end{subequations}
where the expression of $R_k$ is given in Proposition 1; constraint \eqref{C1} ensures different fluid antenna \textcolor{black}{positions} are mutually uncorrelated by enforcing a minimal distance $d_{\min}$; and constraint \eqref{C1} ensures for the user $k$, fluid antenna positions restrict to the feasible region $\mathcal{C}_k$. 

\section{Group Relative Policy Optimization}

In this section, we first perform problem decoupling, then define the MDP, and finally develop a solution using GRPO. 

\subsection{Problem Decoupling}

Since rates for different users are not coupled, we can transform Problem (P1) into $K$ parallel problems. Therefore, we only need to address one of them, given by 
\begin{subequations}
    \begin{eqnarray}
 (\text{P2}) \,\,\,\max_{\{\mathbf{u}_{1},\mathbf{u}_{2}\}} &  \!\!\! \log_2 |\mathbf{I}_2 + P \widetilde{\mathbf{H}} \widetilde{\mathbf{H}}^H \textcolor{black}{{\bf \Omega}}^{-1}|
  \nonumber \\
  \text{s.t.} & \!   \|\mathbf{u}_{n} - \mathbf{u}_{n'} \|_2 \ge d_{\min}, \,\forall  n, n', \label{CC1} \\
  \!\!\!\!\!  & \!\!\!\!\!  \mathbf{u}_{n} \in \mathcal{C}, \qquad\qquad \quad \forall n, \label{CC2} 
\end{eqnarray}
\end{subequations}
where the index $k$ is omitted without loss of generality. Nevertheless, the field response channel model renders Problem (P2) non-convex, making it intractable for standard convex optimization techniques. This challenge motivates the adoption of GRPO DRL algorithm \cite{GRPO}.

\subsection{MDP Definition}
To solve Problem (P2) using DRL, we first formulate a Markov decision process (MDP) characterized by the following key components: 
\begin{itemize}
\item \textbf{State} $\mathbf{s}$: Defined as the set of users' estimated CSI vectors across two fluid antenna positions $\{\widetilde{\mathbf{h}}[1], \widetilde{\mathbf{h}}[2]\}$.
\item \textbf{Action} $\mathbf{a}$: Represented by the set of fluid antenna positions $\{\mathbf{u}_{1}, \mathbf{u}_{2}\}$, which serves as the optimization variable in Problem (P2).

\item \textbf{Reward} $r$: Defined as
\begin{equation}
    r =  \log_2 |\mathbf{I}_2 + P \widetilde{\mathbf{H}} \widetilde{\mathbf{H}}^H \textcolor{black}{{\bf \Omega}}^{-1}|,
\end{equation}
which aligns with the maximization objective of Problem (P2). If any of the constraints \eqref{CC1} or \eqref{CC2} are violated, a large negative reward is assigned.

\item \textbf{Observation} $\mathbf{o}$: Corresponds to the resulting state transition after executing an action $\mathbf{a}$ in state $\mathbf{s}$.
\end{itemize}

\subsection{Proposed GRPO Solution}

DRL operates within the MDP framework, where agents learn optimal policies by maximizing cumulative rewards. This is achieved through state-action value functions approximated by neural networks. GRPO is a non-trivial variant of PPO \cite{PPO}, distinguished by its lower resource consumption during training, as highlighted in \cite{GRPO}. Specifically, GRPO replaces PPO's resource-intensive critic network with a set of trajectories, referred to as a ``group.'' The advantage function for a given policy is computed using relative advantage estimates within this group. This advantage is then utilized to update the actor network via stochastic gradient descent.

According to \cite{GRPO}, GRPO aims to optimize the parameters $w$ by maximizing the objective $J_\text{GRPO}(w)$, which typically comprises a policy gradient term and a regularization term, given in \eqref{GRPO}.
\begin{figure*}
	\begin{equation}
		J_\text{GRPO}(w) = \mathbb{E}\left[\frac{1}{G} \sum_{g=1}^G \frac{1}{O_g} \sum_{t=1}^{O_g} \left\{ 
		\min \left(
		\frac{\pi_w(\textbf{o}_{g,t}|\textbf{a},\textbf{o}_{g,<t})}{\pi_{w_\text{old}}(\textbf{o}_{g,t}|\textbf{a},\textbf{o}_{g,<t})}
		, \text{clip}\left(\frac{\pi_w(\textbf{o}_{g,t}|\textbf{a},\textbf{o}_{g,<t})}{\pi_{w_\text{old}}(\textbf{o}_{g,t}|\textbf{a},\textbf{o}_{g,<t})},1-c,1+c\right)\right)\widehat{A}_{g} - \mu \mathbb{D}_\text{KL}[\pi_{w}\|\pi_\text{ref}]
		\right\}\right] \label{GRPO}
	\end{equation}
	\hrule
\end{figure*}
As its key innovation, the advantage function in \eqref{GRPO} is computed through group relative advantage estimation:
\begin{equation}
	\widehat{A}_{g} = \frac{r_g - \text{mean}(\{r_1,\ldots,r_G\})}{\text{std}(\{r_1,\ldots,r_G\})}, \qquad g \in [G], \label{27} 
\end{equation}
where $G$ denotes the size of the group, the operations $\text{mean}\{\cdot\}$ and $\text{std}\{\cdot\}$ denote the mean and standard deviation of the set $\{\cdot\}$, respectively.

\begin{algorithm}[!t]
	\caption{GRPO Training Procedure}
	\label{algorithm:GRPO}  	
	Obtain the reference policy $\pi_{\text{ref}}$ by training PPO \cite{PPO} for $T$ steps \\
	\For{iteration $=1,2,\ldots,I$}{
		Update old policy: $\pi_{w_{\text{old}}} \leftarrow \pi_w$ \\		 
		Sample $G$ observations $\{\mathbf{o}_g\}_{g=1}^G$ from $\pi_{w_{\text{old}}}(\cdot|\mathbf{s}_1)$, where $\mathbf{s}_1$ is an initial action  \\
		Compute trajectory-wise rewards $\{r_g\}_{g=1}^G$ for each observation $\mathbf{o}_g$ \\
		Estimate advantages $\{\widehat{A}_g\}_{g=1}^G$ using  \eqref{27} \\
		GRPO update \textbf{do} \\
			\qquad Update policy parameters $w$ by maximizing the GRPO objective in \eqref{GRPO}
		
	}
	\BlankLine
	\KwOut{Optimized policy $\pi_{w}$}
\end{algorithm}

  \begin{table}[t]
	\centering
	%\small
	\caption{\textsc{Configuration { \& Hyperparameters}}}
	\label{tab:optimized}
	\begin{tabularx}{0.95\columnwidth}{@{} X X X X @{}}
		\toprule
		\multicolumn{4}{@{}c}{\textbf{Actor Network (GRPO, PPO)}} \\
		\textbf{Component} & \textbf{Type} & \textbf{Input/Output} & \textbf{Activation}\\
		Input Layer       & Linear      & Input Size$\times$256    & ReLU\\ 
		Hidden Layer      & Linear      & 256$\times$256           & ReLU \\     
		Output Layer      & Linear      & 256$\times$Output Size  & --     \\ 
        \multicolumn{4}{@{}c}{\textbf{Critic Network (PPO)}} \\
		\textbf{Component} & \textbf{Type} & \textbf{Input/Output} & \textbf{Activation}\\
		Input Layer       & Linear      & Input Size$\times$256    & ReLU\\ 
		Hidden Layer      & Linear      & 256$\times$256           & ReLU \\     
		Output Layer      & Linear      & 256$\times$Output Size  & --     \\ 
          \multicolumn{4}{@{}c}{{\textbf{Hyperparameters (GRPO, PPO)}}} \\
{\textbf{Learning}} & {\textbf{KL}} & {\textbf{Clipping}} & {\textbf{Batch}} \\
{\textbf{Rate}} & {\textbf{Penalty}} & {\textbf{Threshold}} & {\textbf{Size}} \\
{$0.0001$} & {$0.00001$} & {$0.05$} & {$256$}\\    
\hline
	\end{tabularx} 
	%\vspace{-1in}
\end{table}

The trajectory-wise reward \( r_g \) is set to equal the reward at the terminal observation \( \mathbf{o}_g \). This choice of a final-reward signal is motivated by the structure of our problem, where the solution to Problem (P2) is only available at the end of the trajectory. Furthermore, to ensure stable learning, the clip function in \eqref{GRPO} is introduced following \cite{GRPO} to limit the magnitude of policy updates, which is given by
\begin{eqnarray}
	&& \!\!\!\!\!\!\!\!\!\!\!\! \text{clip}\left(\frac{\pi_w(\textbf{o}_{g,t}|\textbf{a},\textbf{o}_{g,<t})}{\pi_{w_\text{old}}(\textbf{o}_{g,t}|\textbf{a},\textbf{o}_{g,<t})},1-c,1+c\right) \nonumber \\
	&&\!\!\!\!\!\!\!\!\!\!\!\!  = \max\left(\min\left(\frac{\pi_w(\textbf{o}_{g,t}|\textbf{a},\textbf{o}_{g,<t})}{\pi_{w_\text{old}}(\textbf{o}_{g,t}|\textbf{a},\textbf{o}_{g,<t})},1+c\right),1-c\right),
\end{eqnarray}
where $c$ denotes the clipping threshold.

Acting as a regularizer, the Kullback-Leibler (KL) divergence in \eqref{GRPO} is used to keep the updated policy close to the reference policy $\pi_\text{ref}$. The influence of this constraint is determined by the KL penalty coefficient $\mu$. Following \cite{GRPO}, the value of the KL divergence can be approximated by
\begin{eqnarray}
	&&    \mathbb{D}_\text{KL}[\pi_{w}\|\pi_\text{ref}] \approx \frac{\pi_\text{ref}(\textbf{o}_{g,t}|\textbf{a},\textbf{o}_{g,<t})}{\pi_w(\textbf{o}_{g,t}|\textbf{a},\textbf{o}_{g,<t})}  -  \nonumber \\
    && \qquad \qquad \qquad \quad \log \frac{\pi_\text{ref}(\textbf{o}_{g,t}|\textbf{a},\textbf{o}_{g,<t})}{\pi_w(\textbf{o}_{g,t}|\textbf{a},\textbf{o}_{g,<t})} -1.
\end{eqnarray}

Finally, the overall procedure for GRPO training is summarized in Algorithm 1. The algorithm begins by invoking the PPO for a limited number of $T$ steps to obtain a reference policy. It is important to note that $T$ is set to a small value, and the PPO algorithm is not required to converge in this phase.

\begin{figure} 
  \centering
  \includegraphics[width=0.98\linewidth]{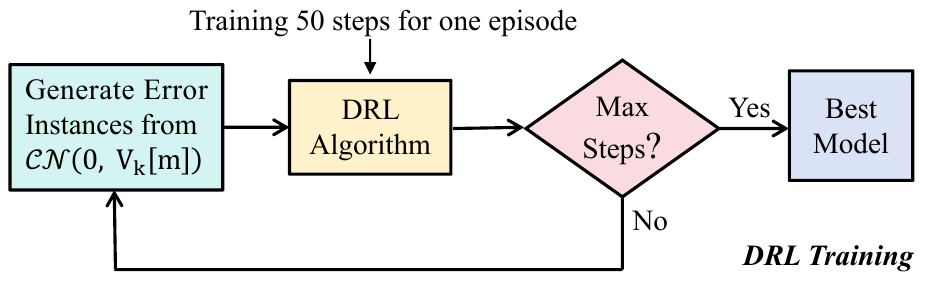}
  \caption{Illustrating the DRL algorithms (PPO-Init, PPO, GRPO) training for our simulations.}
  \label{fig:ppo_power_comparison}
\end{figure}

\section{Simulations}

\subsection{Evaluation Setup}
In simulations, the AWGN variance is set to $\sigma^2 = 10^{-9}$. The system adopts $4$ transmit and receive paths, and serves $K=4$ users, each located $3$m from the BS. The transmit power is set to $0.25$W. The fluid antenna is constrained within a $0.5\text{m}^3$ cubic region, and the carrier frequency is $60$GHz. The covariance matrix is configured as 
\begin{equation}
    \mathbf{V}_k[m] = [0.02, 0.01;  0.01, 0.02] \times \eta,
\end{equation}
for all $k$ and $m$, modeling spatially correlated channel estimation errors commonly encountered in practical systems. All algorithms are for solving Problem (P2). All experiments are conducted on a computing platform equipped with INTEL i7-14700KF CPU and NVIDIA GeForce RTX 4060 8G GPU. The neural network architectures and hyperparameters of PPO and GRPO, the latter employing only a policy network, are detailed in Table I. Fig. \ref{fig:ppo_power_comparison} depicts the DRL training loop, in which error instances are sampled from a complex Gaussian distribution, the policy is updated until $50$ steps, and the optimal model is retained once the maximum step count is reached.
The following baselines are as follows:
 
\textbf{PPO-Init}: We first train the PPO algorithm proposed in \cite{PPO} for $100$K steps to obtain a reference policy for GRPO. This process, which serves as the baseline, requires approximately $2$ minutes on our computing platform.  

\textbf{PPO}: This baseline implements the PPO algorithm as proposed in \cite{PPO} and trains it for $3$ million steps. This baseline implements the widely adopted PPO algorithm \cite{PPO}, which utilizes both actor and critic networks with clipped objective optimization to ensure stable policy updates, thus providing a solid performance benchmark for comparison.

\textbf{MaximumGain}: This heuristic baseline discretizes the feasible region of the fluid antenna into $50000$ slot positions. The algorithm then selects the position yielding the maximum channel gain, measured by the vector 2-norm. The final result is averaged by $1000$ instances.

\textbf{RandomGain}: This heuristic baseline discretizes the feasible region of the fluid antenna into $50000$ slot positions. The algorithm then selects two positions from a uniform distribution. The final result is averaged by $1000$ instances.

\subsection{Results and Discussion on CSI Error Influence}

\begin{figure}[t]
    \centering
    \begin{subfigure}{0.24\textwidth}
        \centering
        \includegraphics[width=\linewidth]{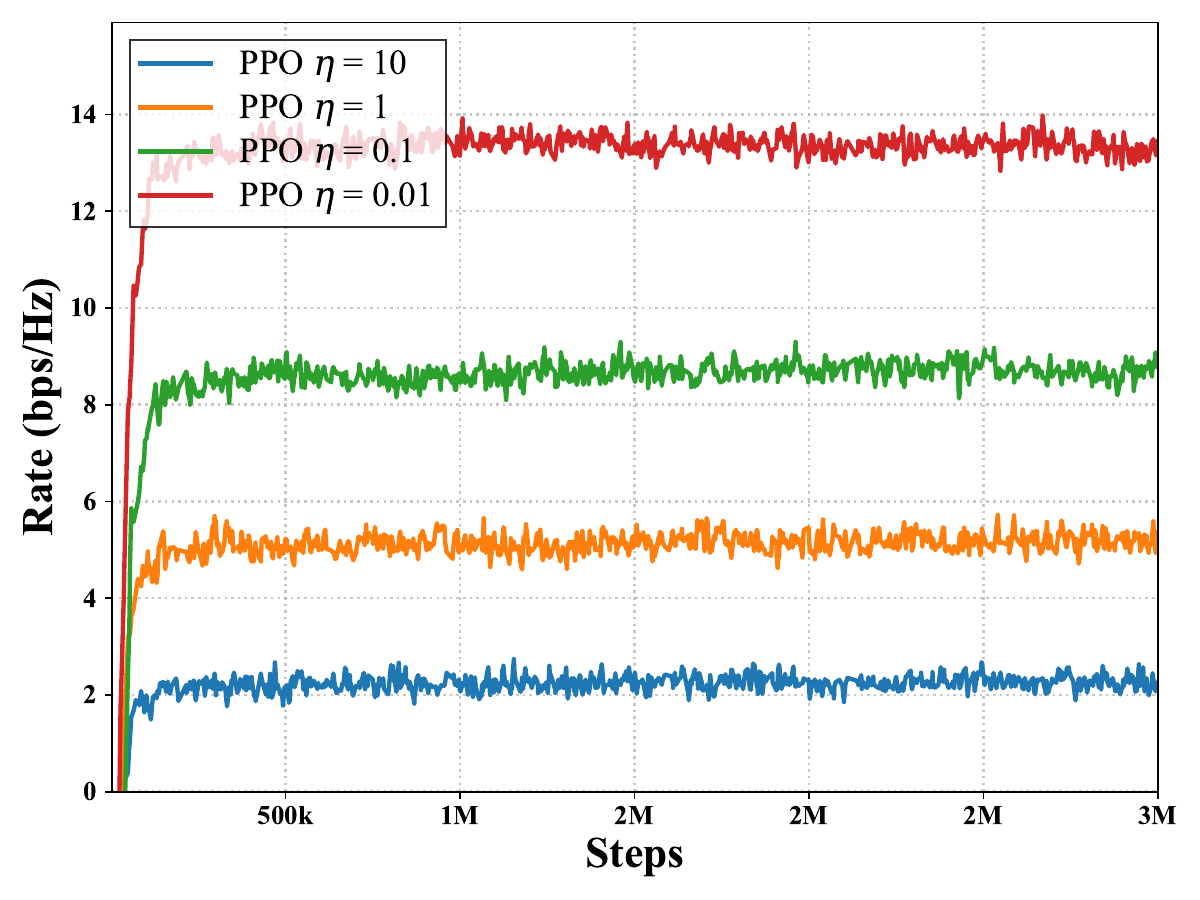}
        \caption{Training}
        \label{F1:a}
    \end{subfigure}%
    \hfill%
    \begin{subfigure}{0.24\textwidth}
        \centering
        \includegraphics[width=\linewidth]{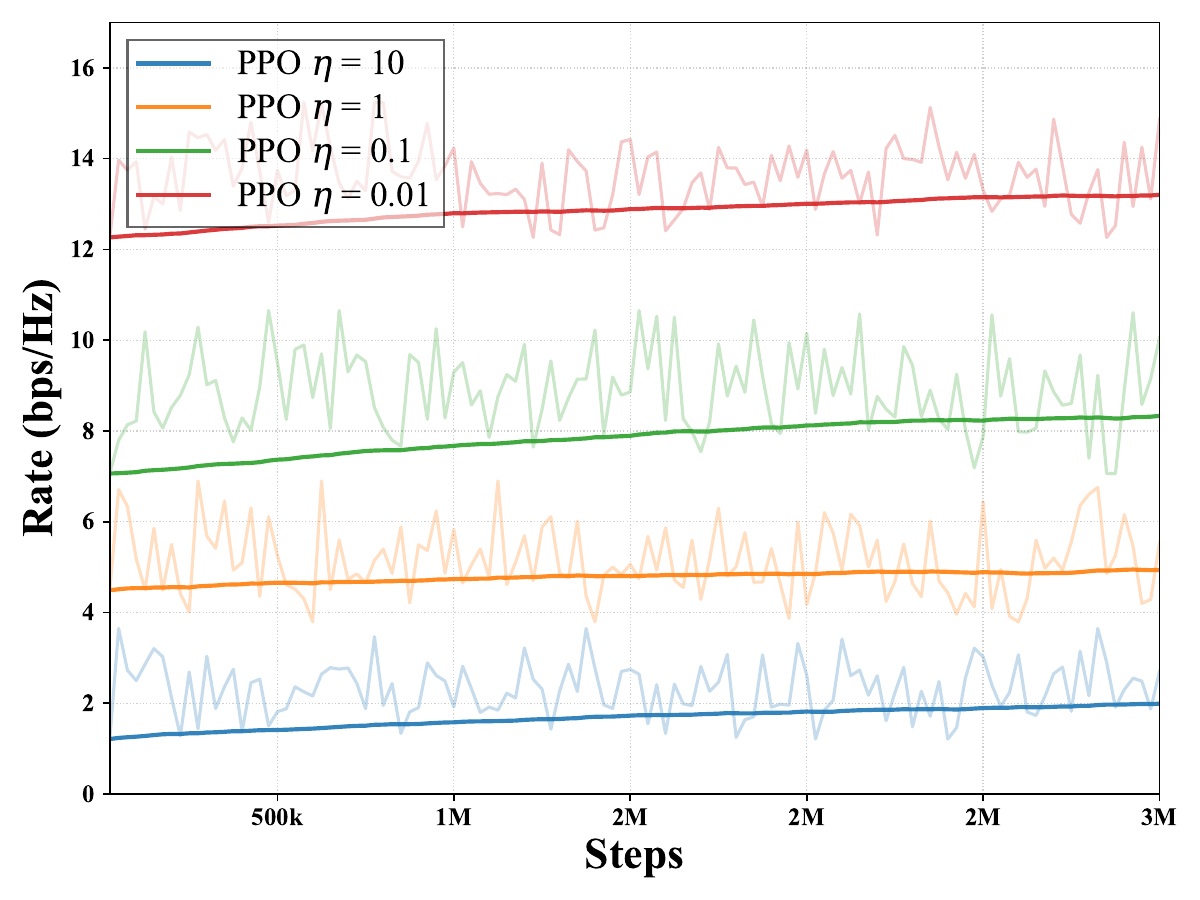}
        \caption{Evaluation}
        \label{F1:b}
    \end{subfigure}%
    \caption{PPO baseline.}
    \label{FF1}
    \end{figure}
    \begin{figure}[t]
  \centering
    \centering
    \begin{subfigure}{0.24\textwidth}
        \centering
        \includegraphics[width=\linewidth]{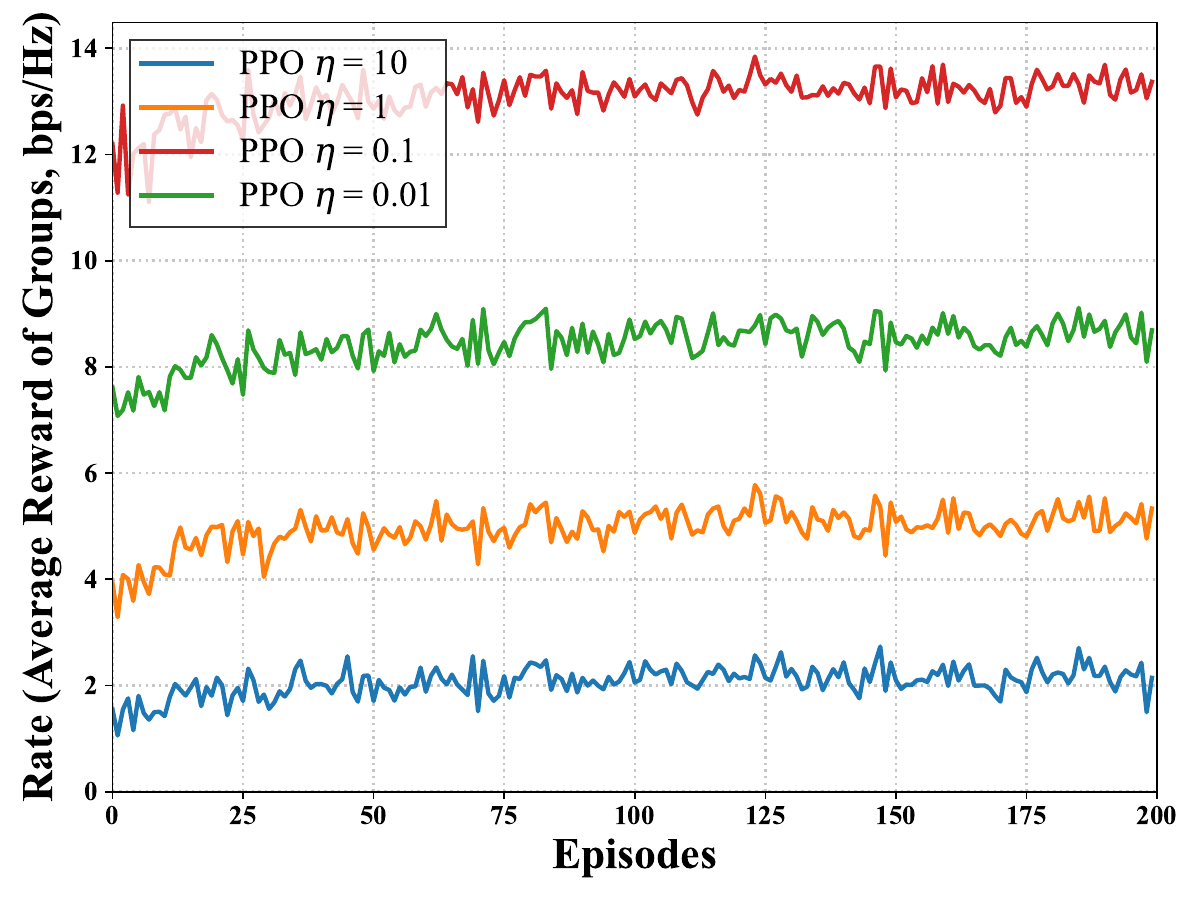}
        \caption{Training}
        \label{F2:a}
    \end{subfigure}%
    \hfill%
    \begin{subfigure}{0.24\textwidth}
        \centering
        \includegraphics[width=\linewidth]{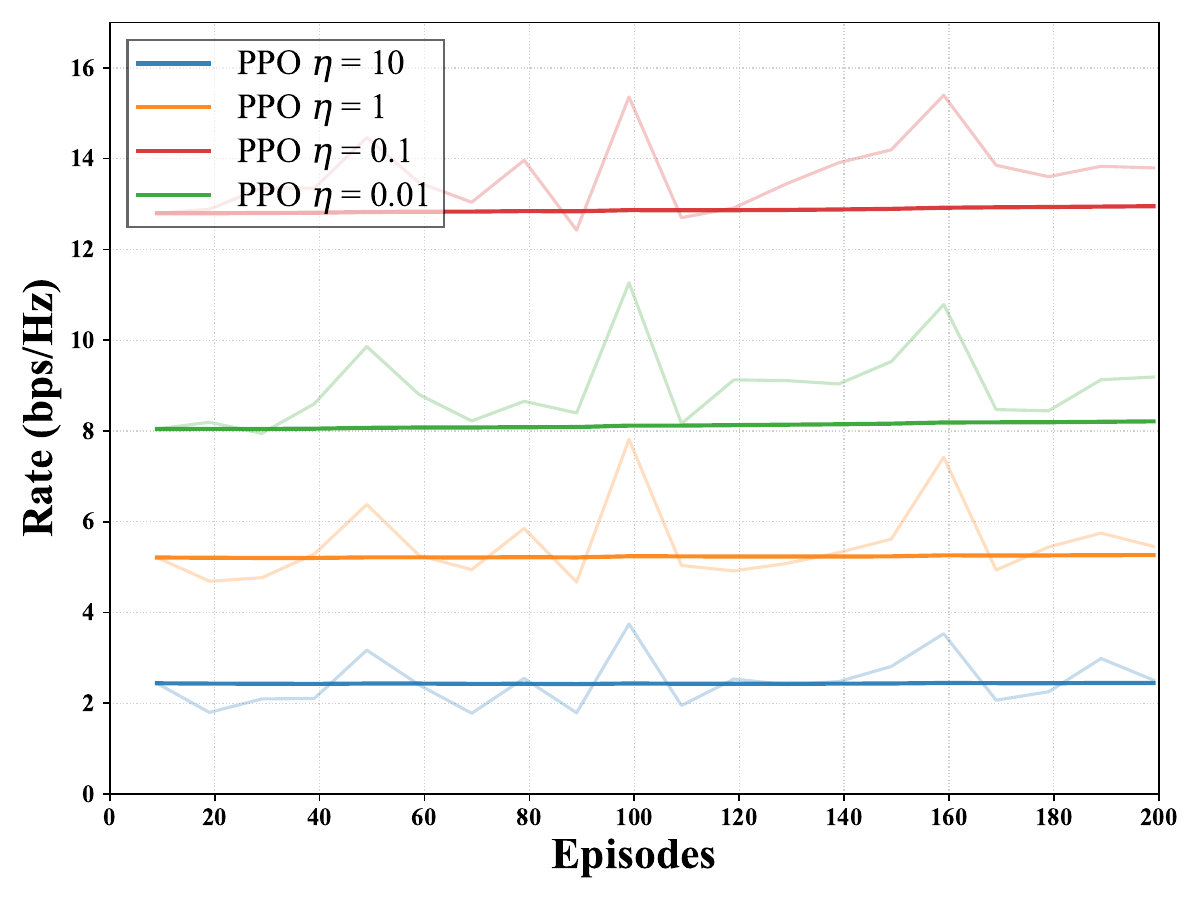}
        \caption{Evaluation}
        \label{F2:b}
    \end{subfigure}%
    \caption{Proposed GRPO solution.}
    \label{FF2}
       \end{figure}

Fig.~\ref{FF1} shows the training and evaluation curves of PPO baseline across different $\eta$, which represents different degree of CSI error. Robustness to CSI error is related to generalization in neural networks. Therefore, we evaluate the PPO baseline in an environment that differs from that in training while maintaining the same error distribution. Fig.~\ref{F1:a} shows that the training curves first increase to a value and oscillate around it. This is because the CSI error is stochastic and the trained policy needs adaptation. To reduce stochastic fluctuations, the evaluation curves in Fig.~\ref{F1:b} are smoothed using an exponential moving average with a factor of $0.99$. The resulting smoothed curves are presented in the same figure. Fig.~\ref{F1:b} shows that the evaluation results are consistent with the training results, indicating effective generalization in the neural networks.

Fig.~\ref{FF2}  shows the training and evaluation curves of proposed GRPO solition across different $\eta$, where the group size and trajectory length are set to $50$ and $50$, respectively. Compared to PPO, the model size and  FLOPs are reduced by $49.6\%$ and $46.7\%$, respectively. This improvement stems from replacing PPO's neural network critic with a simple group-based critic. Fig.~\ref{F2:a} shows that the training curves initially rise and then oscillate around a value. Fig.~\ref{F2:a} shows that the evaluation results are comparable to the training results.  This implies GRPO training is effective. 

\begin{figure}[t]
    \centering
  \includegraphics[width=0.90\linewidth]{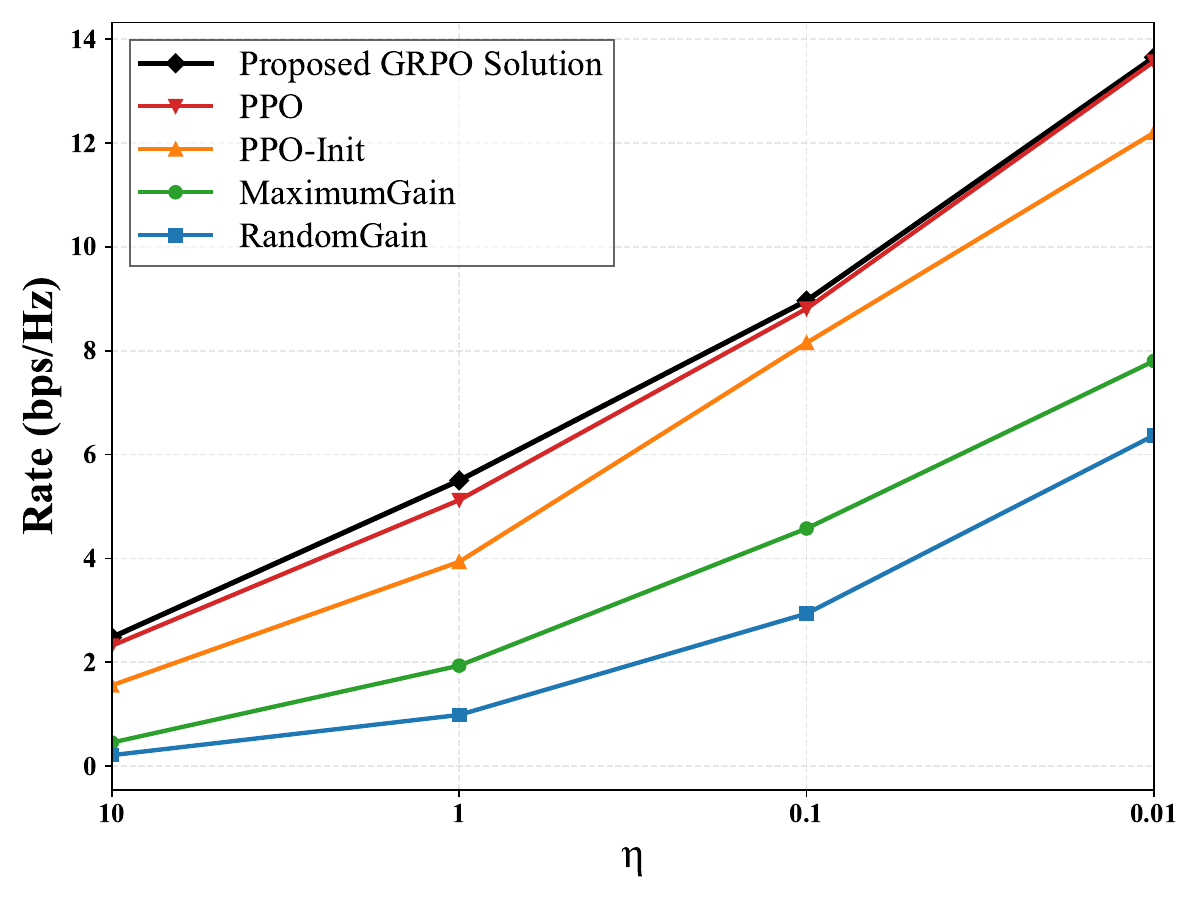}
  \caption{Performance comparison.}
  \label{FF3}
\end{figure}

Fig.~\ref{FF3} compares the proposed GRPO solution with baseline results. Fig.~\ref{FF3} shows the rate grows exponentially with the reduction of CSI error, qualified by $\eta$. This shows that our system is highly sensitive to CSI error, and reducing this error is significantly beneficial. For the severe CSI error, i.e., $\eta = 10$, Fig.~\ref{FF3} shows that heuristic baselines achieve considerably lower rates, whereas DRL algorithms maintain a competent level of performance. Table \ref{FF3-table} validates this observation for $\eta = 10$, showing substantial performance gains of GRPO over heuristic baselines, as well as increased gains over PPO and PPO-Init. This result demonstrates the effectiveness of GRPO against CSI error, and other PPO baselines.
In average gains over different $\eta$ in Table \ref{FF3-table}, GRPO achieves  $4.17\%$, $30.29\%$, $200.78\%$, $465.38\%$ over PPO, PPO-Init, MaximumGain, and  RandomGain, respectively. 

\begin{table}[t]
  \centering
  \caption{GRPO gains over baselines (\%)}
  \label{FF3-table}
  \begin{tabular}{lrrrr}
    \toprule
    Baseline & $\eta = 10$ & $\eta = 1$ & $\eta = 0.1$ & $\eta = 0.01$ \\
    \midrule
    PPO & 6.95 & 7.40 & 1.74 & 0.59 \\
    PPO-Init & 59.42 & 39.86 & 9.97 & 11.89 \\
    MaximumGain & 447.77 & 184.47 & 95.91 & 74.97 \\
    RandomGain & 1083.41 & 458.52 & 205.28 & 114.30 \\
    \bottomrule
  \end{tabular}
\end{table}

\section{Conclusion}
This paper presented a novel fluid antenna-driven BIA framework that achieves robust performance without CSIT. Our main contributions include: 1) the first integration of FAS with BIA under imperfect CSI, 2) a robust sum-rate maximization formulation via fluid antenna positioning, and 3) the GRPO solution that halves the model size and FLOPs of PPO while enhancing performance through group-based exploration. Simulations verified the following gains: $4.17\%$ over PPO, $30.29\%$  over PPO-Init, and $200.78\%$ and $465.38\%$ over heuristic MaximumGain and RandomGain. Our study showed that GRPO leverages the generalization capability of neural networks to learn error distributions and employs group-based exploration that resets the ``critic", thereby achieving excellent robustness in stochastic environments.

\begin{appendices}
    \section{Proof of Proposition 1}
 First, we need to derive the covariance matrix of estimation error and AWGN.   
 \begin{eqnarray}
    &&  \textcolor{black}{{\bf \Omega}} = P{\bf \Delta}_k \mathbb{E}\{\mathbf{s}_k \mathbf{s}_k^H\} {\bf \Delta}_k^H + 
     \mathbb{E}\{\mathbf{n}_k \mathbf{n}_k^H\} \nonumber \\
    &&  \quad\,   =  P{\bf \Delta}_k {\bf \Delta}_k^H + \begin{bmatrix}
         K \sigma^2 & 0 \\
         0 & \sigma^2
     \end{bmatrix}  \nonumber \\
   && \!\!\!\!\!\!\!\!   = P\begin{bmatrix} 
           {\bf{\Delta}}_k[1] {\bf{\Delta}}_k^H[1]  & {\bf{\Delta}}_k[1] {\bf{\Delta}}_k^H[2] \\
     {\bf{\Delta}}_k[2] {\bf{\Delta}}_k^H[1] & {\bf{\Delta}}_k[2] {\bf{\Delta}}_k^H[2]  
      \end{bmatrix} + \begin{bmatrix}
         K \sigma^2 & 0 \\
         0 & \sigma^2
     \end{bmatrix} = \, \eqref{Rn}. \nonumber  
 \end{eqnarray}
 Next, we substitute the expression of $\textcolor{black}{{\bf \Omega}}$ into the MIMO capacity formula in \cite[Eqn. 2]{Qingjiang}, leading to the desired result. 
\end{appendices}

\bibliographystyle{ieeetr} 
\bibliography{ICCFAS}

\end{document}